\documentclass[12pt]{article}
\usepackage{graphicx}
\title{Electronic structure of negatively curved graphene}
\author{D.\,V.\,Kolesnikov$^{+*}$,
V.\,A.\,Osipov$^+$ \\ e-mail: kolesnik@theor.jinr.ru,\,
osipov@theor.jinr.ru$^+$,\\ $^*$Joint Institute for Nuclear
Research,\\ Bogoliubov Laboratory of Theoretical Physics,\\ 141980
Dubna, Moscow region, Russia}

\begin{document}

 \maketitle\begin{abstract}We study the electronic structure of graphene in the
presence of either sevenfolds or eightfolds by using a gauge
field-theory model. The graphene sheet with topological defects is
considered as a negative cone surface with infinite Gaussian
curvature at the center. The density of electronic states is
calculated for a single seven- and eightfold as well as for a pair
of sevenfolds with different morphology. The density of states at
the Fermi energy is found to be zero in all cases except two
sevenfolds with translational factor $M\neq 0$.\end{abstract}
  Nowadays graphene and related graphene-based nanostructures are
the subject of great interest. It is well known that some number
of fivefolds must be included in the honeycomb graphene lattice to
produce various nanostructures with positive Gaussian curvature of
the surface: nanocones, nanohorns, closed nanotubes, fullerenes,
etc. There is another type of possible defects, sevenfolds and
eightfolds, which results in a surface with negative Gaussian
curvature. Experimentally the presence of sevenfolds in various
nanostructures was observed by Iijima \cite{iijima92}. The
structures with sevenfolds only are rather exotic~\cite{terrones},
however, the presence of fivefold-sevenfold pairs is expected in
some cases and, what is important, leads to new features of the
graphene-based materials. For example, in the carbon nanotube this
pair changes chirality, which allows one to connect metallic and
semiconducting nanotubes without breaking of $\pi$-bonds (see,
e.g., \cite{eletskii}). Therefore the question of the electronic
structure of carbon nanostructures containing sevenfolds has the
definite theoretical and practical interest.

Earlier computations on the carbon lattice~\cite{tamura} shows the
general decrease of the density of states (DoS) near the
sevenfolds in the sixfold plane. Notice that the influence of the
curvature was not taken into account in~\cite{tamura}. Effects of
both topological defects and local curvature on the electronic
properties of planar graphene has been recently studied
in~\cite{vozmediano1,vozmediano2} within the effective low-energy
field-theory model proposed to study the electronic structure of
slightly curved graphene sheets with an arbitrary number of
pentagon-heptagon pairs.

In the present paper, we investigate the electronic structure of
the graphene plane with either a single sevenfold or two
sevenfolds/one eightfold inserted (which corresponds to the
negative 60$^\circ$ and 120$^\circ$ disclination, respectively).
The surface of the plane is assumed to be free to bend but
impossible to stretch. This results in appearance of the
``surplus-angle cone" surface with infinite (negative) Gaussian
curvature in the center, but with absence of the rotational
symmetry. We use a field-theory model suggested in
~\cite{osipov_math,okp,jetpl} and, in final form, in~\cite{rjp},
which includes two gauge fields to take into account both the
presence of sevenfolds and the influence of surface curvature.

Let us consider the planar honeycomb lattice, which can be bent
but can not be stretched. A single sevenfold can be inserted in
this lattice by using the standard ``cut-and-paste" procedure. The
surface of the lattice due to the presence of negative
disclination will appear to be curved (see Fig. 1). Let us assume
the curved surface in the form
\begin{eqnarray}
  \vec r=(r\cos\varphi\sin\theta(\varphi),
  r\sin\varphi\sin\theta(\varphi),r\cos\theta(\varphi)),
\label{surfc}
\end{eqnarray}
where
$$\theta(\varphi)=\frac{\alpha}{2}\cos(2\varphi)+\frac{\pi}{2}.$$
\begin{figure}[ht]
\begin{center}
\includegraphics[width=8.5cm]{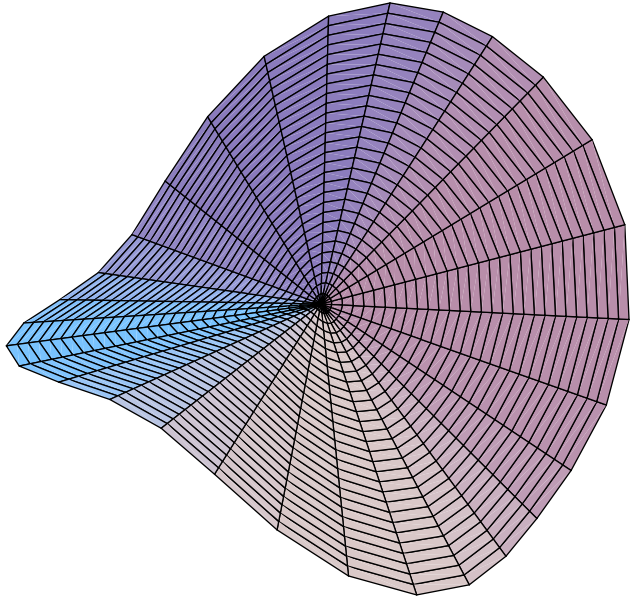}
 \caption[]{The curved surface of the honeycomb lattice with a single
sevenfold inserted at the center (a ``surplus-angle
cone").}\label{lat1}
\end{center}
\end{figure}
After the cut-and-paste procedure, the length of the line of a
constant distance $R$ from the sevenfold (which is also the line
of intersection of the 3d-sphere and the surface) is larger than
for the same line in the planar case. Since the sevenfold leads to
an additional $60^\circ$ sector, the length should have the value
$(7/6)2\pi R$. One can define it from the condition
\begin{eqnarray}
  R \int_0^{2\pi}
  \sqrt{\sin^2 \theta(\varphi)+(\theta'(\varphi))^2}d\varphi=2\pi
  R\frac{7}{6},
\label{llength}
\end{eqnarray}
which allows us to fix the only parameter $\alpha$. The metrical
tensor for the described surface is written as
\begin{eqnarray}
  g_{rr}=1,g_{\varphi\varphi}=r^2[\sin^2\theta(\varphi)+\alpha^2 \sin^2(
  2\varphi)],g_{r\varphi}=0, \label{surfg}
\end{eqnarray}
and the non-zero metrical connection coefficients read
\begin{eqnarray}
  \Gamma^r_{\varphi\varphi}=-g_{\varphi\varphi}/r,\Gamma^\varphi_{\varphi
  r}=\Gamma^\varphi_{r\varphi}=1/r,\Gamma^\varphi_{\varphi\varphi}=(1/2)\partial_\varphi\ln
  g_{\varphi\varphi}.
\label{gammas}
\end{eqnarray}
The two-folds can be choosen in the form
\begin{eqnarray}
  e^1_r=1,e^2_\varphi=\sqrt{g_{\varphi\varphi}},e^1_\varphi=e^2_r=0.
\label{es}
\end{eqnarray}
By using the known relation $(\omega_\mu)^{ik}=e^i_\nu D_\mu
e^{\beta\nu}$ where an ordinary covariant derivative $D_\mu$
includes the metrical connection (\ref{gammas}), one finds the
spin connection coefficients as
\begin{eqnarray}
  (\omega_\varphi)^{12}=-(\omega_\varphi)^{21}=-\frac{g_{\varphi\varphi}}{r}.
\label{omegas}
\end{eqnarray}
The model Dirac equation has the form (see~\cite{rjp})
\begin{eqnarray}
-i\sigma^a e_a^\mu(\nabla_\mu
-ia_\mu^k-iW_\mu)\psi^k=E\psi^k,\label{2dcur}
\end{eqnarray}
where $E$ is measured from the Fermi energy, $\sigma^a\, (a=1,2)$
are the conventional Pauli matrices, $a_\mu^k\ (k=K,K_-)$ and
$W_\mu$ are two gauge fields (non-Abelian and Abelian,
respectively), and $\nabla_\mu=\partial_\mu+\Omega_\mu\,
(\mu=r,\varphi$) with
$$\Omega_\mu=\frac{1}{8}{\omega^a_\mu}^b
[\sigma_a,\sigma_b]\label{bomega}.$$ In our case, one obtains
$$
\Omega_\varphi=-i\sigma_3\frac{\sqrt{g_{\varphi\varphi}}}{2r},\,\Omega_r=0.
$$
To simplify the problem the gauge fields are taken in the same
form as for the isotropic case:
\begin{equation}a_\varphi^k=\pm
1/4,\, W_\varphi=-1/6,a^k_r=W_r=0\label{aw}.\end{equation} The
Dirac equation (\ref{2dcur}) can be diagonalized in respect to the
$K/K_-$- part (see~\cite{rjp} for detail), so that finally it
takes the form
\begin{eqnarray}
  -i\sigma_1(\partial_r+\frac{1}{2r})\psi-\frac{i\sigma_2}{\sqrt{g_{\varphi\varphi}}}
  (\partial_\varphi-ia_\varphi-iW_\varphi)\psi=E\psi.
\label{7dirac}
\end{eqnarray}

One can easily see that for $\alpha=0$ (\ref{7dirac}) reduces to
the planar Dirac equation in the polar coordinates. Unlike the
planar case, in (\ref{7dirac}) $g_{\varphi\varphi}$ is a function
of the polar angle $\varphi$. This markedly complicates the
analysis. Let us make the separation of variables in
(\ref{7dirac}) assuming
$$\psi(r,\varphi)=\left(\begin{array}{c}
  u(r) \\
  v(r)
\end{array}\right) \Phi(\varphi).$$
The function $\Phi$ should neglect the dependence of the
wavefunction on $\varphi$ in (\ref{7dirac}). Therefore, we find
the equation for $\Phi$ in the form
\begin{eqnarray}
  \frac{\Phi'}{\Phi}-ia_\varphi-iW_\varphi=ij\sqrt{\sin^2\theta(\varphi)+\alpha^2 \sin^2(
  2\varphi)
  },
\label{Feq}
\end{eqnarray}
where $j$ is a constant. Introducing an auxiliary function
\begin{eqnarray}
  G(\varphi)=\int_0^\varphi\sqrt{\sin^2\theta(\phi)+\alpha^2 \sin^2(
  2\phi)
}d\phi, \label{Gfun}
\end{eqnarray}
one can write down $\Phi$ in the form
$\Phi=\exp(ia_\varphi\varphi+iW_\varphi\varphi)\exp(ijG(\varphi))$.
Let us note that $G(2\pi)=2\pi (7/6)$ due to (\ref{llength}). The
boundary condition for the spinor wavefunction has the form
$\psi(\varphi+2\pi)=-\psi(\varphi)$. This allows us to find the
constant $j$ as
\begin{eqnarray}
  j=j_n=\frac{6}{7}[1/2+(n-a_\varphi-W_\varphi)],\, n=0,\pm
  1,\ldots
\label{jn}
\end{eqnarray}
and
\begin{eqnarray}
  \Phi_n=\exp(i(a_{\varphi}+W_{\varphi})\varphi)\frac{\exp(ij_n
  G(\varphi))}{\sqrt{2\pi}},
\label{Phin}
\end{eqnarray}
 which finally leads to the system of coupled
equations for $u_n,v_n$ in the following form:
\begin{eqnarray}
  -i(\partial_r
  +\frac{1}{2r}+\frac{j_n}{r})v_n(r)=Eu_n(r),\nonumber\\
 -i(\partial_r
  +\frac{1}{2r}-\frac{j_n}{r})u_n(r)=Ev_n(r).
\label{uveq1}
\end{eqnarray}
Notice that this system is equivalent to the planar Dirac equation
with the only difference in the values of $j_n$. The solution to
(\ref{uveq1}) is written as
\begin{eqnarray}
  u_n=C  J_{|j_n-1/2|}(|E|r), \; v_n=\pm iC  J_{|j_n+1/2|}(|E|r),
\label{uvs}
\end{eqnarray}
where $J$ are the Bessel functions, and C is the normalization
constant. The density of electronic states per unit area (LDoS)
can be found from the properly normalized wavefunctions. The
normalizing condition
$$
 C^2\int r( u^2+v^2) dr=1.
$$
allows us to find the normalization constant. Taking into account
the asymptotical behaviour of the solution (\ref{uvs})
($[J_{n+1/2}^2(x)+J_{n-1/2}^2(x)]x\simeq 2/\pi$ when
$x\rightarrow\infty$) one can find the LDoS in the following form:
\begin{eqnarray}
  LDoS(E,r)= \frac{|E|}{2\pi(1+\Omega)} \; \sum_{\epsilon,\,n}
  (J_{\epsilon(j_n+1/2)}(|E|r)^2
+ J_{\epsilon(j_n-1/2)}(|E|r)^2). \label{dos1}
\end{eqnarray}
Here $\Omega$ ($\Omega=1/6$ for a sevenfold) defines a surplus
angle of a cone and
\begin{equation}\label{j7}
  j_n=6/7(1/2+n+1/6\pm 1/4).
\end{equation}
The summation runs over all values of n and $\epsilon=\pm 1$,
which satisfies the condition $\epsilon j_n\geq 0$. This
restriction comes from the fact that the LDoS should be finite (or
zero) in the limit ${E\rightarrow 0}$. In that way it overrides
the normalization condition for the wavefunction at $r\rightarrow
0$. Notice that the LDoS does not depend on the angle coordinate
$\varphi$ despite the fact that the surface does not possess the
axial symmetry. Accordingly, (\ref{dos1}) formally looks like the
LDoS for a normal (deficit-angle) cone found in \cite{crespi}. As
a result, the only dimensional quantity in the LDoS is $r$ while
the energy can be reduced to the dimensionless parameter with
respect to the energy unit $\hbar V_F/r$ where $V_F$ is the Fermi
velocity.

In the case of two closely situated sevenfolds or one eightfold
the LDoS is also described by (\ref{dos1}) with $\Omega=1/3$ and
\begin{equation}\label{j8}
j_n=6/8(1/2+n+1/3\pm(1/2+M/3)).
\end{equation}
The factor $M$ describes the morphology of defects: $M=0$ for an
eightfold or when the shift vector $(n,m)$ from one sevenfold to
another satisfies the condition $(n+m)\; mod\; 3=0$, otherwise
$M=1$ (see \cite{rjp,crespi} for detail). The energy dependence of
the LDoS near the center and far from the center is presented in
Fig. 2.
\begin{figure}[ht]
\begin{center}
\includegraphics[width=7.5cm]{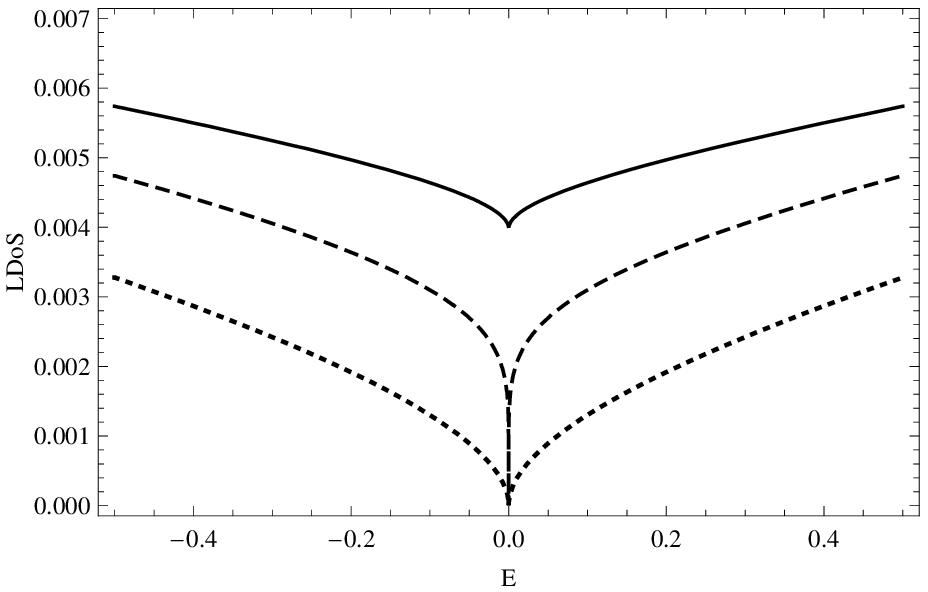}\\
\includegraphics[width=7.5cm]{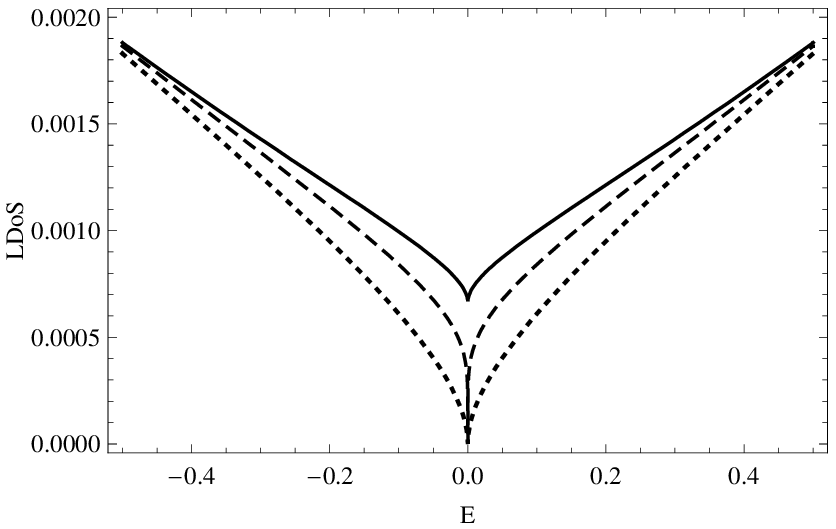}
\caption[]{Fig.2. The local density of states per unit area (in
the same scale), as a function of the energy near the center (top)
and far from the center (bottom). }\label{dose}
\end{center}
\end{figure}
As is seen, the LDoS rapidly increases with energy near the
sevenfold while far from the defect it has almost linear behavior.

The LDoS as a function of the energy and the coordinate is shown
in Fig. 3.
\begin{figure}[ht]
\begin{center}
\includegraphics[width=7.5cm]{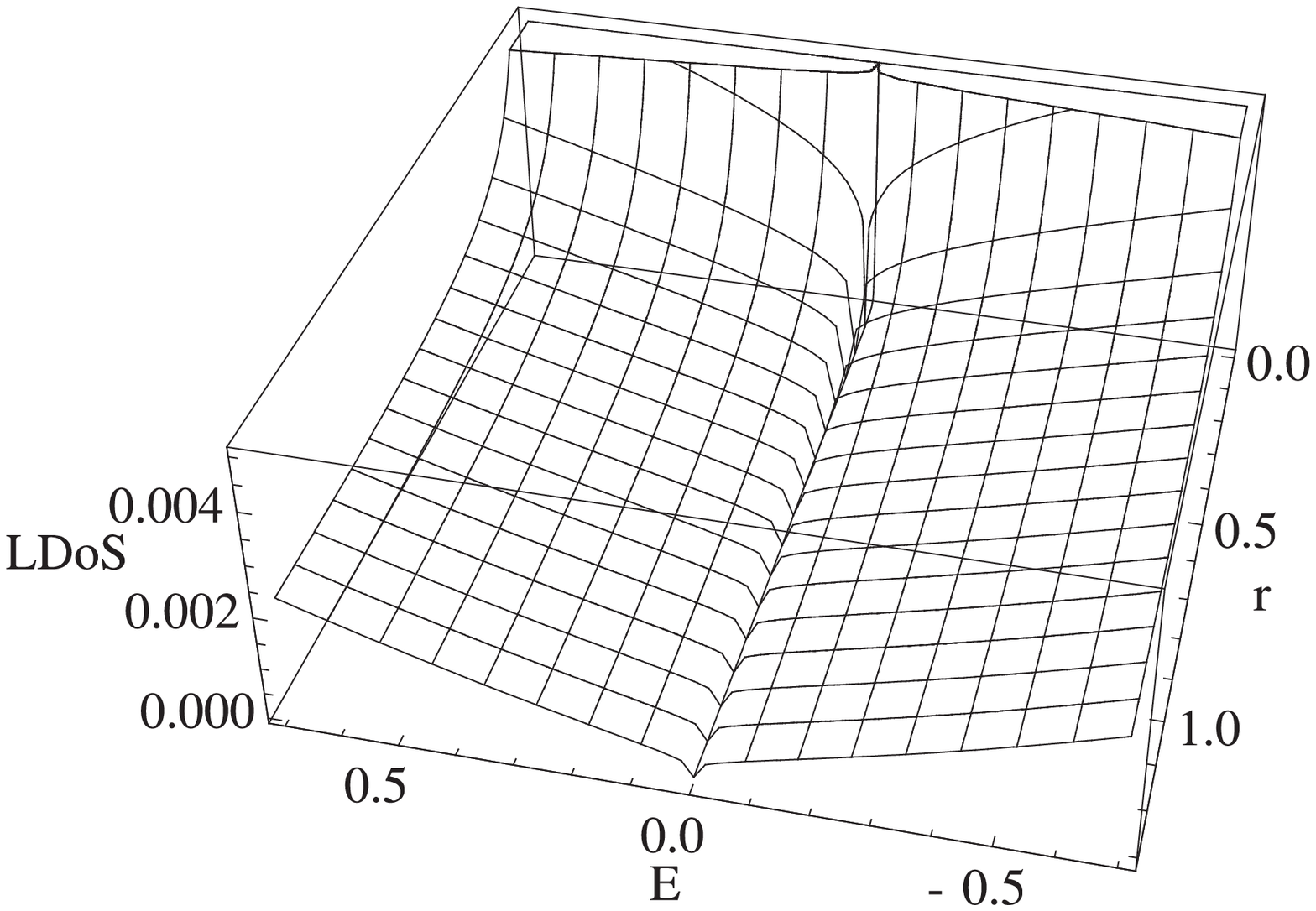}\\
\includegraphics[width=7.5cm]{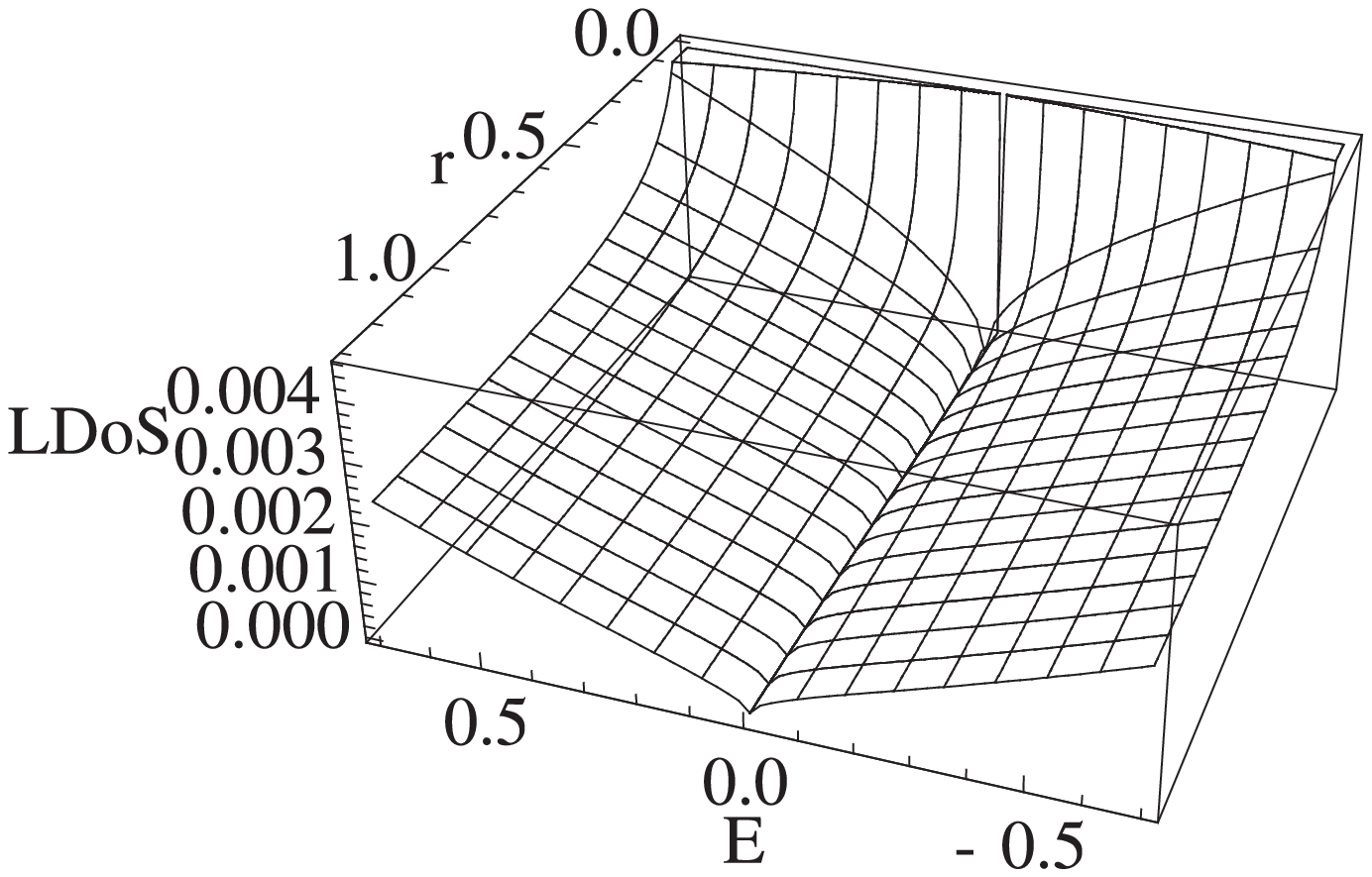}\\
\includegraphics[width=7.5cm]{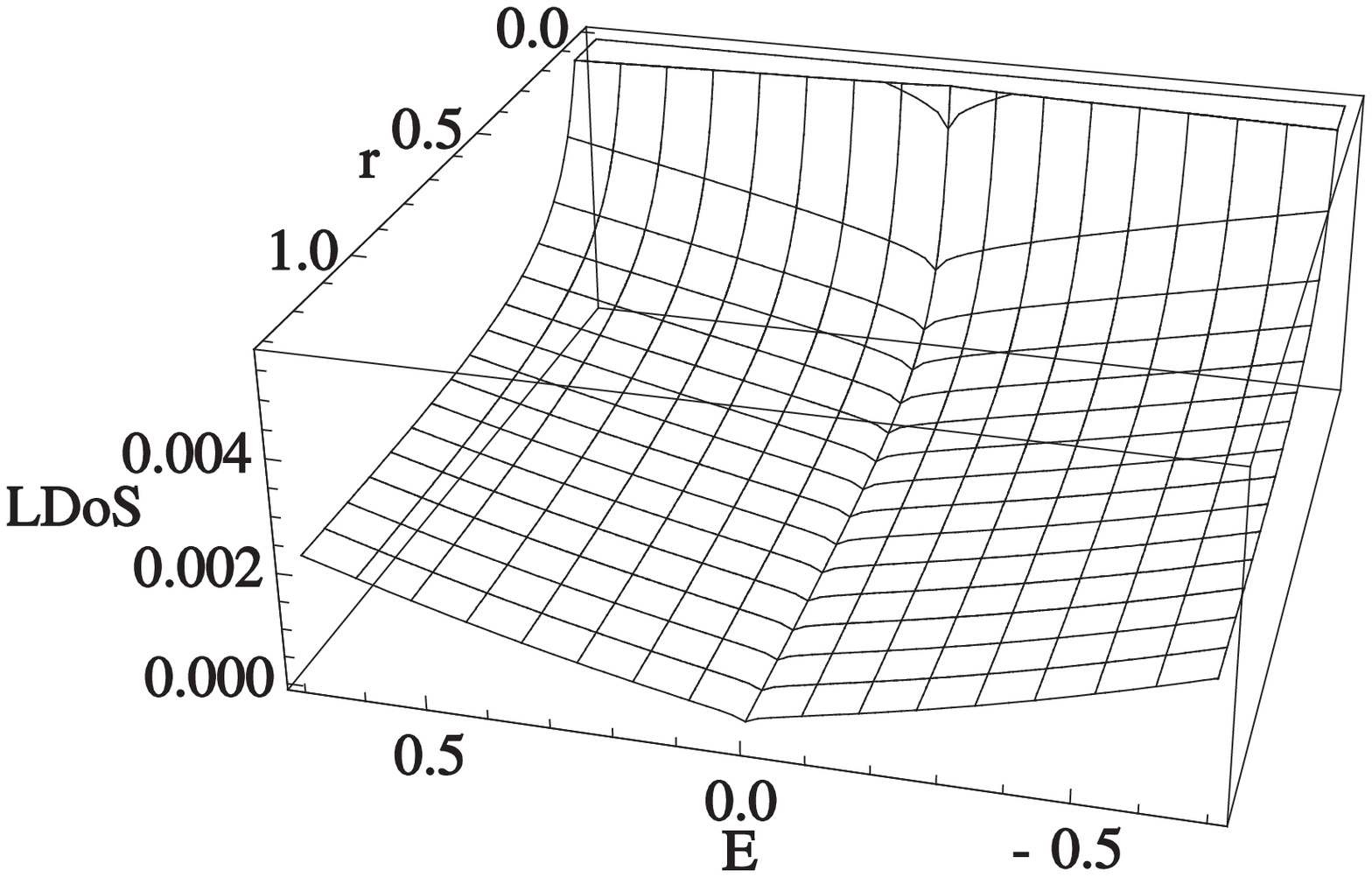}\\
 \caption[]{The local density of states per unit area (in the same
scale), as a function of the radial coordinate r and the energy E.
The cases of a single sevenfold (top), single eightfold or two
sevenfolds with $M=0$ (middle), and two sevenfolds with non-zero
$M$ (bottom) are shown. }\label{dos3d}
\end{center}
\end{figure}
One can see that at the Fermi energy ($E=0$) the LDoS is zero for
both single defects while a finite value of LDoS is found for two
sevenfolds with $M\neq 0$. This is provided by the term with
$j_n=0$ in the LDoS. The LDoS increases near the center ($r\approx
0$) and becomes r-independent at far distances being linear in
energy like in the case of planar graphene.

To conclude, we have investigated the electronic structure of the
graphene with one and two sevenfolds as well as with one eightfold
in the framework of the field-theory approach. The curvature of
the surface was taken into account by considering the "negative
cone" with infinite and negative Gaussian curvature at the center.
The calculations show that near defects the density of states is
growing faster in comparison with the planar case. When the energy
and/or a distance from the defect increases, the LDoS tends to be
almost similar to the planar case. These findings are in general
agreement with numerical calculations in \cite{tamura}. At small
distances from the defect an increase of the LDoS (per unit area)
was found. Within our approach this increase can be explained by
the influence of infinite curvature at the center ($r=0$). In the
case of two sevenfolds with the translational factor $M\neq 0$ the
existence of the metallization near the defects was observed. For
the eightfold, the metallization in the defect region is absent.
This conclusion differs from the results of \cite{tamura} where
near the eightfold a non-zero value of the LDoS was found to vary
(highly anisotropic) for different atoms on the same distance from
the defect, thus forming local metal states. Such difference can
be explained by two reasons: (a) the continual character of our
model and (b) an isotropic approximation. Notice that in our study
the LDoS does not depend on the angle $\varphi$ due to excluded
radial components of the fields (\ref{aw}). On the other hand, the
shape of the surface is propertly taken into account in our model,
leading to curvature-induced peculiarities which are absent in the
planar case~\cite{tamura}.

This work has been supported by the Russian Foundation for Basic
Research under grant No 08-02-01027.

\end{document}